\documentclass[%
 aip,
 amsmath,amssymb,
reprint,%
]{revtex4-1}

\usepackage{graphicx}
\usepackage{dcolumn}
\usepackage{bm}

\usepackage[utf8]{inputenc}
\usepackage[T1]{fontenc}
\usepackage{mathptmx}
\usepackage{etoolbox}
\usepackage{comment}
\usepackage{makecell}
\usepackage{float}
\usepackage{bbm}

\makeatletter
\def\@email#1#2{%
 \endgroup
 \patchcmd{\titleblock@produce}
  {\frontmatter@RRAPformat}
  {\frontmatter@RRAPformat{\produce@RRAP{*#1\href{mailto:#2}{#2}}}\frontmatter@RRAPformat}
  {}{}
}%
\makeatother
\begin{document}

\preprint{AIP/123-QED}

\title[FIBID Pt]{Rubidium Focused Ion Beam Induced Platinum Deposition}
\author{Y. Li}
\author{S.Xu}
\affiliation{%
Department of Applied Physics, Eindhoven University of Technology\\P.O. Box 513, 5600 MB Eindhoven, the Netherlands\\}%
\author{M. Sezen}
\author{F. Bakan Misirlioglu}
\affiliation{Sabanci University Nanotechnology Research and Application Center (SUNUM), 34956, Istanbul, Turkey}
\author{E. J. D. Vredenbregt}
 \email{e.j.d.vredenbregt@tue.nl}
\affiliation{%
Department of Applied Physics, Eindhoven University of Technology\\P.O. Box 513, 5600 MB Eindhoven, the Netherlands\\}%

\date{\today}

\begin{abstract}
This work presents characterization of focused ion beam induced deposition (FIBID) of platinum using both rubidium and gallium ions. Trimethylplatinum [$\mathrm{(MeCp)Pt(Me)_3}$] was used as the deposition precursor. Under similar beam energies, 8.5 keV for $\mathrm{Rb^+}$ and 8.0 keV for $\mathrm{Ga^+}$, and beam current, near 10 pA, the two ion species deposited Pt films at 0.90 $\mathrm{\mu m^3/nC}$ and 0.73 $\mathrm{\mu m^3/nC}$ respectively. Energy-dispersive x-ray spectroscopy shows that the $\mathrm{Rb^+}$ FIBID-Pt consists of similar Pt contents (49\% for $\mathrm{Rb^+}$ FIBID and 37\% for $\mathrm{Ga^+}$ FIBID) with much lower primary ion contents (5\% Rb and 27\% Ga) than the $\mathrm{Ga^+}$ FIBID-Pt. The deposited material was also measured to have a resistivity of $8.1\times 10^4$ $\mathrm{\mu\Omega\cdot cm}$ for the $\mathrm{Rb^+}$ FIBID-Pt and $5.7\times 10^3$ $\mathrm{\mu\Omega\cdot cm}$ for the $\mathrm{Ga^+}$ FIBID-Pt. 
\end{abstract}

\maketitle

\section{\label{sec:introduction}Introduction}

Focused ion beams (FIBs) have become essential tools for material science studies and the semiconductor industry because of the ability to perform ion-microscopy and to precisely remove material on a nanoscale. One widely used FIB function is FIB induced deposition (FIBID), which allows material to be deposited over selected target areas. During FIBID, a precursor gas is introduced to the ion scanning area \cite{utke2008FEIBID}. The chemical bonds of the precursor break within this area and then a film of the desired material is left on the sample substrate. Because of the precise control over the deposition area, FIBID enables procedures such as integrated circuit editing \cite{orloff2009handbook}, photomask repair \cite{robinson1989maskrepair}, transmission electron microscope (TEM) sample preparation \cite{gianuzzi1999FIBreview}, and also 3D nano-structure fabrication \cite{reyntjens2000FIBID3D}. In the case of TEM sample making, protective caps of Pt or W are usually deposited over the region of interest to preserve the surface features. The various applications of FIBID-Pt make it important to understand the properties of the process.

The deposition rate, Pt purity, and resistivity are among the most commonly reported values in articles covering FIBID-Pt \cite{puretz1992FIBIDPtfilm}. The deposition rate describes how fast the Pt deposit grows during the FIBID process. Because of the nature of FIBID, impurity elements from the precursor gas, the FIB chamber, and the incident ions become incorporated into the deposit \cite{telari2002Ptcharacterization}. The resulting less-than-pure Pt deposit leads to a higher resistivity than that of bulk pure Pt. The Pt purity and deposition resistivity are critical for determining the application of the FIBID-Pt.     

As an industrial standard, Ga FIBID-Pt has been studied mostly for high beam energy ($\approx$ 30 keV) and high beam current ($>$ 50 pA) \cite{puretz1992FIBIDPtfilm}\textsuperscript{,} \cite{ttao1990FIBIDPt}. Data from lower energy/current FIBID-Pt are scarce. FIBID using other ion species than $\mathrm{Ga^+}$ can also be of great interest since different applications might require different deposition yield and metal purity. The $\mathrm{Xe^+}$ source has been proved capable of depositing Pt films with properties comparable to the ones created by the Ga LMIS \cite{rue2012XeFIBID}. As an alternative to the Ga LMIS, Cs or Rb have been considered as ion source species \cite{tenhaaf2014ultracoldRBFIB}\textsuperscript{,} \cite{steele2017CsFIB}. These two elements can be laser-cooled to form high-brightness ion beams. In this paper, low energy Rb and Ga FIBID of Pt were investigated for comparison across different primary ion species and beam energies. 

\section{\label{sec:experiment}Experiment}

\subsection{\label{sec:Rb FIB}Rb FIB system}

\begin{figure}
    \centering
    \includegraphics[width=0.5\linewidth]{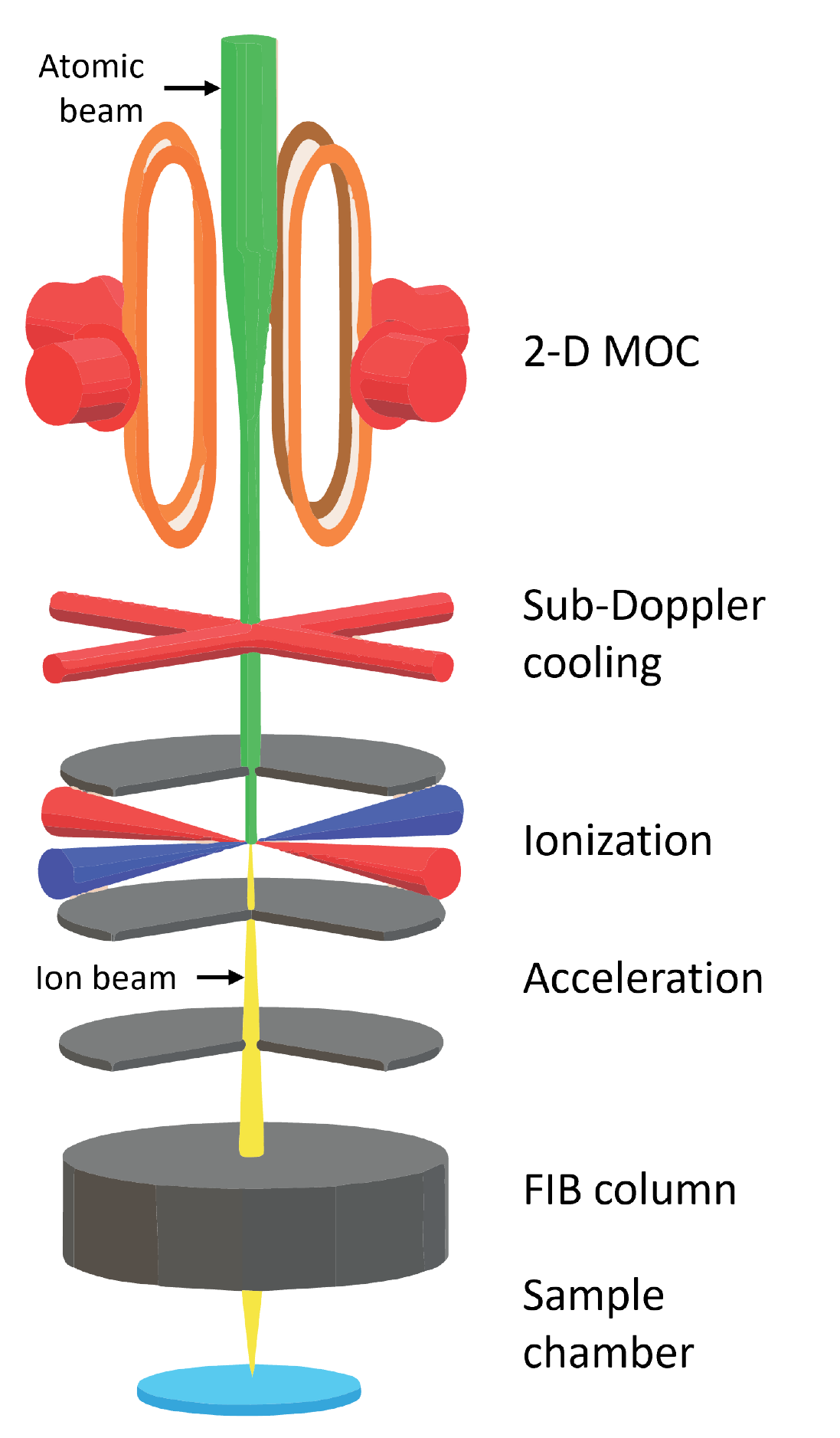}
    \caption{Schematic of the Rb FIB (not to scale), with the ion beam starting at the atomic beam-laser overlap point in the ionization region. The current selection aperture, which is not shown in the diagram, is inside the FIB column.}
    \label{fig:Rb FIB schematic}
\end{figure}

Fig.\ref{fig:Rb FIB schematic} shows the schematic diagram of the Rb FIB used for the deposition work. It mainly incorporates two sections, a laser-cooled $\mathrm{Rb^+}$ ion source\cite{tenhaaf2014ultracoldRBFIB} on top of an FEI FIB200TEM column. Neutral Rb atoms are first heated to $160^\circ$C in a Knudsen cell and then collimated in a heated thin tube. The heating provides the atoms with the velocity needed for propagation along the beam path. After the collimation tube, the atomic beam is cooled and compressed transversely in a 2-D magneto-optical compressor (MOC) under Doppler cooling using a 780-nm laser. Sub-Doppler cooling right outside of this MOC further reduces the transverse temperature of the particles to 40 $\mu$K before they enter the ionization region.

A two-step photoionization approach is used for the Rb FIB. As shown in the ionization region of Fig.\ref{fig:Rb FIB schematic}, two sets of laser beams travel perpendicularly with the intersection overlapping with the Rb atomic beam. A 780-nm laser excites the Rb atoms to an intermediate state and an optical-resonator-augmented 480-nm laser ionizes the excited atoms. This photoionization process happens between two closely placed metal plates. The top plate is kept at a high electric potential, usually at 8.5 keV for most applications. The bottom plate is kept at a few hundred volts lower than the top plate. The potential difference provides a small electric field to draw the ions down the beam path while keeping the energy difference among the ions small. The ions receive further acceleration between the bottom plate and a grounded plate as shown in the acceleration region of Fig.\ref{fig:Rb FIB schematic}. After the acceleration, the ions are focused by the FIB lenses onto the sample in the FIB chamber. 

The $\mathrm{Rb^+}$ ion beam used for this paper was selected by a 79-$\mu$m diameter aperture located in the FEI FIB column. This is to maintain a round beam spot for ease of operation. Therefore, the $\mathrm{Rb^+}$ current was small (usually $<$ 10 pA) compared to some of the FIBID processes discussed in the reference literature\cite{telari2002Ptcharacterization}\textsuperscript{,}\cite{ttao1990FIBIDPt}. The beam was measured to have a diameter of 160 nm (d$_{50}$) and estimated to have an energy spread of 0.8 eV. 

\subsection{\label{subsec:FIBID on Si}FIBID of Pt on Si}

\begin{figure}
    \centering
    \includegraphics[width=0.9\linewidth]{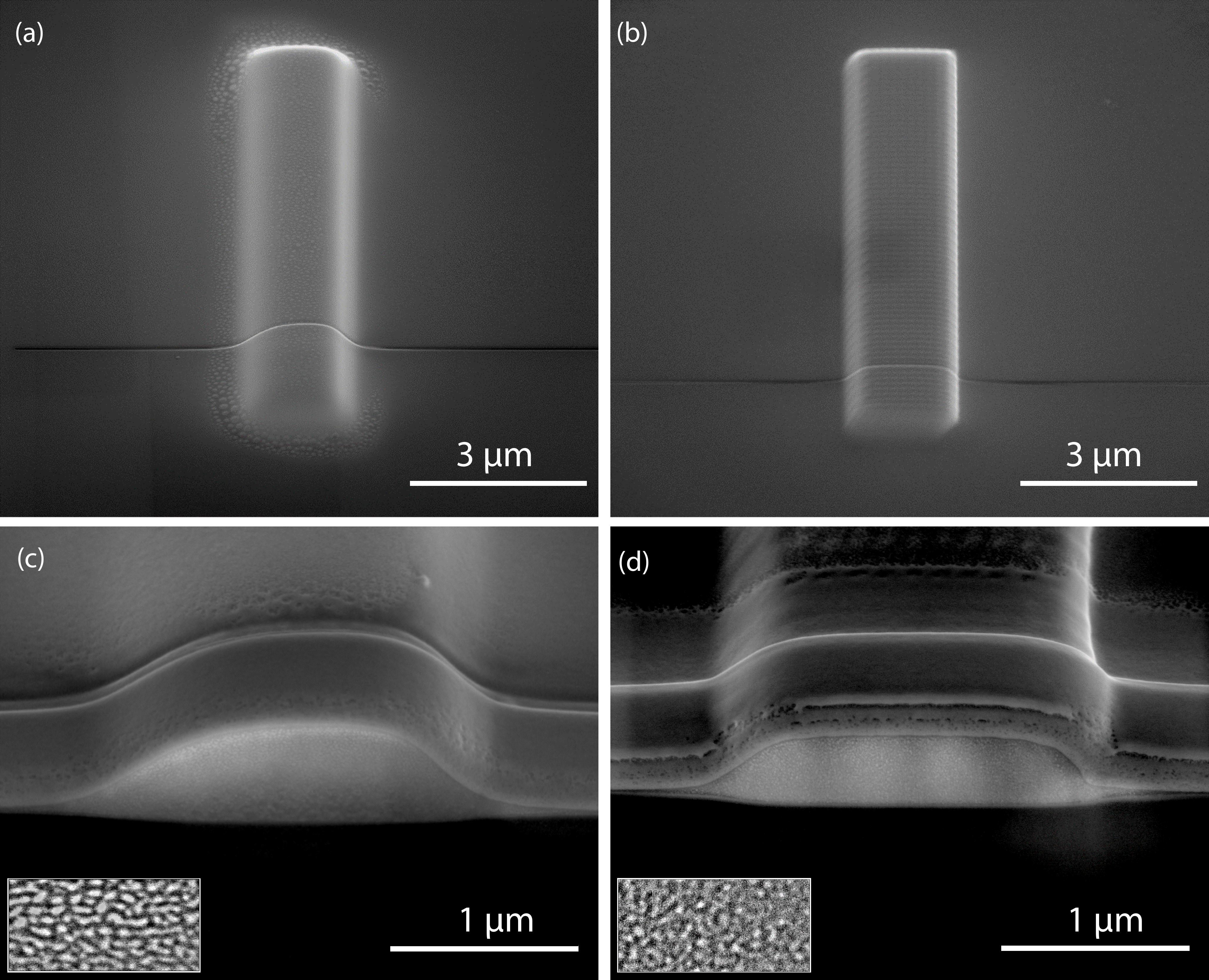}
    \caption{SEM images of $\mathrm{1.5 \times 10 \:\mu m^2}$ FIBID Pt on Si surface by (a) 6.5 pA $\mathrm{Rb^+}$ for 33.75 minutes. (b) 8.5 pA $\mathrm{Ga^+}$ for 25.80 minutes. Grooves were etched across the Pt deposits to indicate volumes. (c) cross section of the $\mathrm{Rb^+}$ FIBID Pt shown in (a). (d) cross section of the $\mathrm{Ga^+}$ FIBID Pt shown in (b). The insets in (c) and (d) contain images of $255\times 122.5\:\mathrm{nm^2}$ areas of the cross section for microstructure analysis.}
    \label{fig:FIBID Pt }
\end{figure}

The Rb FIB discussed in Section \ref{sec:Rb FIB} and a Thermo Fisher Scientific Nova 600i $\mathrm{NanoLab^{TM}}$ DualBeam were used for the Pt deposition. Both systems are equipped with standard Gas Injection System (GIS) hardware. The GIS delivers the precursor gas, Trimethylplatinum [$\mathrm{(MeCp)Pt(Me)_3}$], to the vicinity of the ion-sample interaction area via a nozzle with a 500 $\mu$m diameter. The precursor chemical was heated to 40 \textdegree C to release the gas flow while the sample substrates were kept at room temperature during the deposition process. The exit of the nozzle was maintained about 100 $\mu$m above the sample surface. The Rb FIB was operated at 8.5 keV and the Ga FIB at 8.0 keV beam energy with currents near 10 pA to make the results comparable. Fig.\ref{fig:FIBID Pt }(a) and (b) show the deposition results of a Rb$^+$ and a Ga$^+$ FIBID Pt. 

The deposits were later imaged by the DualBeam scanning electron microscope (SEM) and cross sectioned by the DualBeam Ga FIB. Lamellae were also prepared as TEM samples on the DualBeam for composition analysis via energy-dispersive x-ray spectroscopy (EDS). These EDS measurements were performed in a Thermo Fisher Scientific $\mathrm{Titan^{TM}}$ G2 60-300 TEM equipped with a Super-X G2 detector.

\subsection{\label{sec:resistivity measurements}Resistivity measurements}

\begin{figure}
    \centering
    \includegraphics[width=0.9\linewidth]{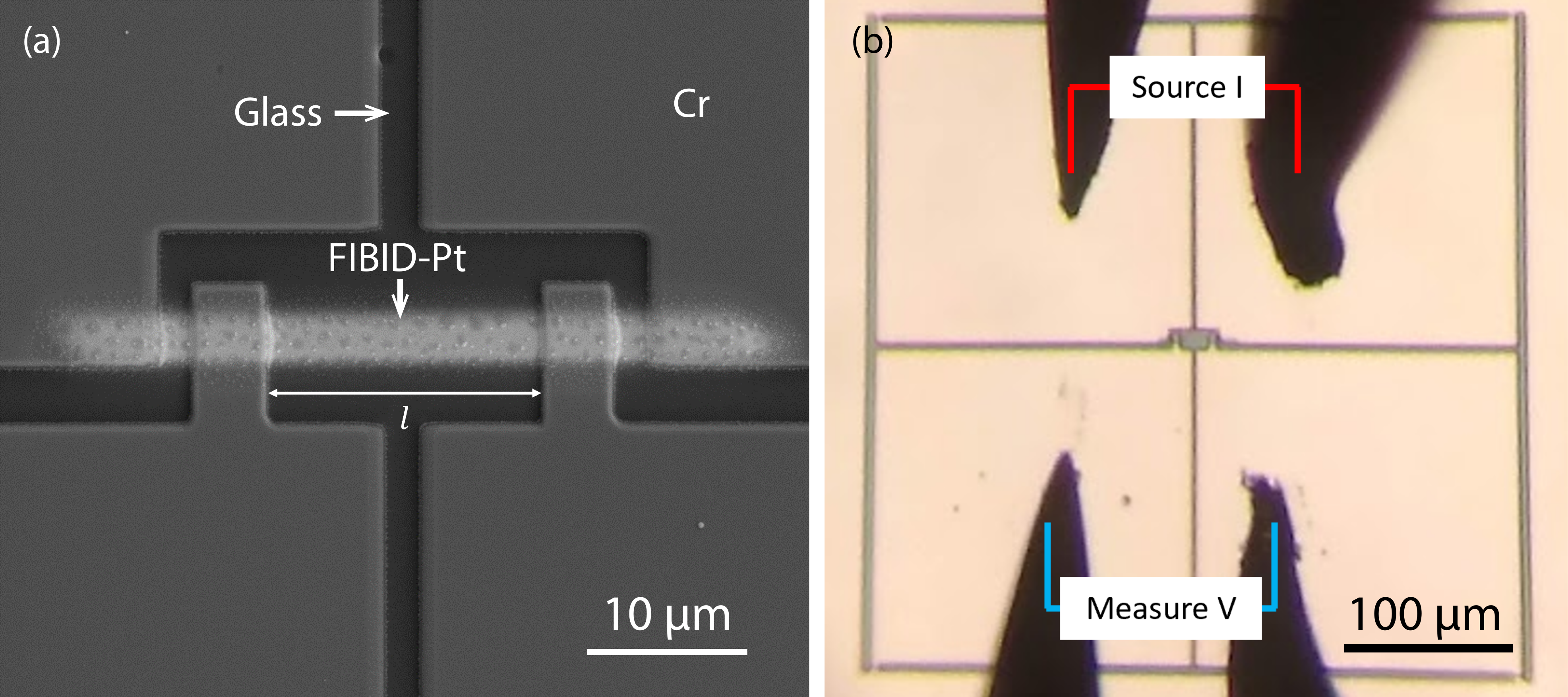}
    \caption{$1.5\times45$ $\mu$m$^2$ Rb$^+$ FIBID Pt for resistivity measurement (a) 10 pA $\mathrm{Rb^+}$ for 65.65 minutes. The effective span of the Pt deposition is $l=17.4$ $\mu$m (b) The same sample viewed on the probe station.}
    \label{fig:resistivity }
\end{figure}

Deposit resistivity was measured on samples with the design illustrated in Fig.\ref{fig:resistivity }(a). This design uses a 4-point probe method by forcing a known current through the deposit while sensing the correlated voltage drop. For each resistivity measurement, a testing structure was first etched on an 80-nm-chromium-on-glass standard by the Ga FIB. The four electrodes of the testing structure were next connected by a FIBID Pt bar. The three distinct gray scale regions in Fig.\ref{fig:resistivity }(a), from low to high intensity, correspond to glass, Cr, and $\mathrm{Rb^+}$ FIBID Pt. After the FIBID, these samples were cleared of deposition over-spray and isolated electrically by the $\mathrm{Ga^+}$ FIB so the current path was only through the Pt deposit.

The prepared FIBID Pt samples were then placed onto a standard probe station with connection illustrated in Fig.\ref{fig:resistivity }(b). The deposit resistance ($R$) could be extracted through linear fitting of the current-voltage data measured by a Keithley 2400 source meter. The Pt deposits were then cross-sectioned after the resistance measurement to achieve the cross-sectional area ($A$). The resistivity ($\rho$) of the deposited material could be calculated using the equation $\rho = R \cdot A \cdot l^{-1}$ for a known span of the deposit ($l$).    

\section{\label{sec:results and discussion}Results and discussion}

\subsection{\label{sec:deposition-rate}Deposition rate}

The deposition rate $\mathbbm{R}_{dep}$ could be calculated by $\mathbbm{R}_{dep} = V \cdot (I \cdot t)^{-1}$, where $V$ is the volume of the deposition, $I$ is the beam current used for the deposition and $t$ is the process time. The Pt deposit bars on Si shown in Fig.\ref{fig:FIBID Pt }(a) and (b) were used for the deposition rate measurement. The $\mathrm{Rb^+}$ current was 6.5 pA and the $\mathrm{Ga^+}$ current was 8.5 pA. The chamber pressure increase due to the flow of the precursor gas was from a base pressure of $<1 \times 10^{-7}$ mbar to $2.3 \times 10^{-6}$ mbar for the $\mathrm{Rb^+}$ FIBID and from $1.3 \times 10^{-6}$ mbar to $1.2 \times 10^{-5}$ mbar for the $\mathrm{Ga^+}$ FIBID. Cross sectional areas were then determined for volume calculations (Fig.\ref{fig:FIBID Pt }(c) and (d)) by ImageJ. The deposition rate for $\mathrm{Rb^+}$ is $0.90_{\pm0.02} \; \mathrm{\mu m^3/nC}$ and for $\mathrm{Ga^+}$ is $0.73_{\pm0.02} \; \mathrm{\mu m^3/nC}$. The visual determination of the deposit volume and the fluctuation in the ion current contribute to the deposition rate uncertainties.

Since $\mathbbm{R}_{dep}$'s were measured directly on the final FIBID product, they represent the net effects of simultaneous deposition and sputtering during the process. To differentiate these two process, net deposition yield $Y_N$ [deposited metal atoms/incident ion] is introduced. $Y_N$ is defined as the difference of the pure deposition yield of the ion beam for a given precursor chemical ($Y_D$) and the sputtering yield of the incident ion on the deposited material ($Y_S$) \cite{jro1993FIBIDAu}. For quantification purposes, $Y_N$'s are given by equating the deposited material to pure Pt of the same volume \cite{ttao1990FIBIDPt}. The measured $\mathbbm{R}_{dep}$'s are equivalent to $Y_N = 9.5$ (Pt/ion) for $\mathrm{Rb^+}$ and $Y_N = 7.7$ (Pt/ion) for $\mathrm{Ga^+}$ under the used beam conditions. 

The $Y_N$ of Pt measured for $\mathrm{Ga^+}$ has a higher value than the literature $Y_N$ \cite{ttao1990FIBIDPt} of a 35 keV 20 pA $\mathrm{Ga^+}$ beam [ranging between 1-1.5 (Pt/ion)]. One potential contribution to this difference is the lower beam energy used in our application. As Chen et al. pointed out \cite{pchen2009SEinFIBID}, $Y_D$ remained almost the same while $Y_S$ decreased by 20\% when beam energy for a $\mathrm{Ga^+}$ FIBID process was reduced from 30 keV to 5 keV. This decrease in $Y_S$ leads to a larger $Y_N$ for low beam energy FIBID. 

\subsection{\label{sec:composition}Deposit composition and microstructure}

EDS measurements were performed on the sample lamellae in a TEM and the results were analysed in FEI's ES Vision software. The results show that the $\mathrm{Rb^+}$ FIBID Pt consists of C:O:Pt:Rb = 25:20:49:5 and the $\mathrm{Ga^+}$ FIBID Pt of C:O:Pt:Ga = 22:14:37:27 in atomic percentage. These results contain 3-5\% statistical uncertainties, which are the typical errors associated with EDS measurements \cite{manual-bruker}. The 27\% Ga content of the Pt deposit is within the Ga\% range (18-30\%) reported in \cite{ttao1990FIBIDPt} for FIBID Pt by 32 keV 20-80 pA $\mathrm{Ga^+}$. The higher vapor pressure of Rb than that of Ga \cite{rhonig1957vaporpressures} might be why the Rb content is lower than the Ga content in the FIBID Pt. Since Rb is more volatile than Ga, it is more likely to leave the deposited material. The decrease in primary ion content of the $\mathrm{Rb^+}$ FIBID Pt gives room to higher at.\% of C, O, and Pt. 

The microstructure of the FIBID Pt was studied using the insets of Fig.\ref{fig:FIBID Pt }(c) and (d). These images reveal that both deposits have the typical features of the FIBID Pt described by Langford et al. \cite{langford2007Ptresistivity}, with Pt-rich grains embedded in a C-rich matrix. The Pt grains appear brighter than the C matrix in the SEM images, thus allowing the grains to be counted by image processing software ImageJ based on a given grey scale threshold. IamgeJ also provides the area of each grain counted. By equating each grain area to a circular disk area, an effective grain diameter could be calculated. The average grain diameter is $13_{\pm3}$ nm for the $\mathrm{Rb^+}$ FIBID Pt and $14_{\pm2}$ nm for the $\mathrm{Ga^+}$ FIBID Pt. 

\subsection{\label{sec:resistivity}Deposit resistivity}

\begin{figure*}
    \centering
    \includegraphics[width=0.9\linewidth]{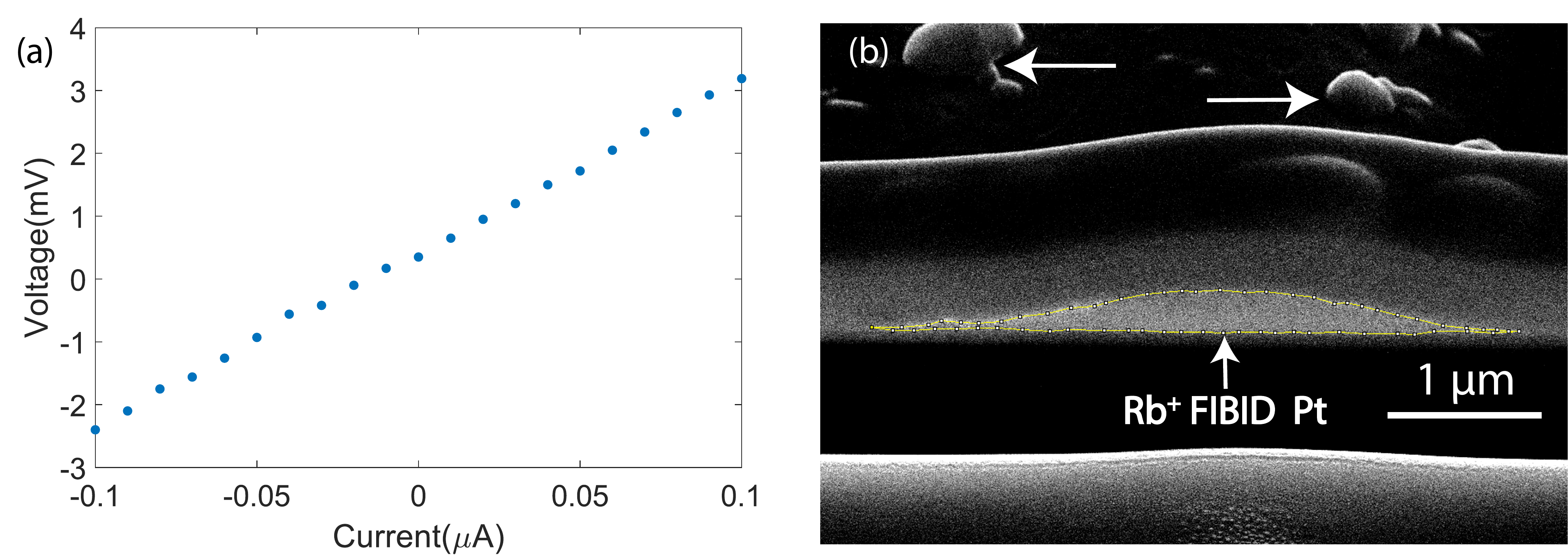}
    \caption{Resistivity measurement of a $\mathrm{Rb^+}$ FIBID Pt (a) I-V data collected on the probe station (b) Cross section with the $\mathrm{Rb^+}$ FIBID Pt circled by the yellow line. The arrows point to the deposit surface bubbles. }
    \label{fig:resistivity measurement}
\end{figure*}

The resistivity of the $\mathrm{Rb^+}$ FIBID Pt is $1.2_{\pm0.4}\times 10^5$ $\mathrm{\mu\Omega\cdot cm}$ and of the $\mathrm{Ga^+}$ FIBID Pt is $6.4_{\pm0.7}\times 10^3$ $\mathrm{\mu\Omega\cdot cm}$. The measurements were performed on two deposits made by 10 pA $\mathrm{Rb^+}$ (one sample shown in Fig.\ref{fig:resistivity }(a)) and two by 7.8 pA $\mathrm{Ga^+}$. The ion doses of the deposition were kept at $1.6\times10^{18}$ ions/cm$^2$. The total patterning area for each deposition was defined as $1.5\times45$ $\mu$m$^2$ while the effective span of the resistance measurement $l$ was measured between the inner edges of the Cr electrodes. This is because the resistance of the FIBID Pt was assumed to be much higher than that of the Cr. Fig.\ref{fig:resistivity measurement}(a) shows the I-V data collected on the $\mathrm{Rb^+}$ FIBID Pt and Fig.\ref{fig:resistivity measurement}(b) contains the cross section of the deposit after the probe station measurement. The dominating uncertainties came from the resistance difference among the samples, which is about 31\% for the $\mathrm{Rb^+}$ FIBID Pt and 12\% for the $\mathrm{Ga^+}$ FIBID Pt.     

 The resistivity reported in literature for $\mathrm{Ga^+}$ FIBID Pt \cite{utke2008FEIBID}\textsuperscript{,}\cite{ttao1990FIBIDPt}\textsuperscript{,} \cite{rue2012XeFIBID}\textsuperscript{,}\cite{langford2007Ptresistivity} falls in a large range of $10^3$-$10^5$ $\mathrm{\mu\Omega\cdot cm}$. Among these values, the resistivity of FIBID Pt by 30 keV 30 pA $\mathrm{Ga^+}$ was measured to be $1\times10^3$-$5\times10^3$ $\mathrm{\mu\Omega\cdot cm}$. The resistivity of our $\mathrm{Ga^+}$ FIBID Pt measurements is close to the upper limit $5\times10^3 \mathrm{\mu\Omega\cdot cm}$, which gives credence to our resistivity measurement procedure. The smaller beam current used leads to the slightly higher resistivity of our measurement than the literature value, as Tao et al. pointed out that the deposit resistivity decreases as the beam current increases \cite{ttao1990FIBIDPt}. 

The higher O\% and C\% of the $\mathrm{Rb^+}$ FIBID Pt contribute to its higher resistivity than that of the $\mathrm{Ga^+}$ FIBID Pt. Both Tao et al.\cite{ttao1990FIBIDPt} and Telari et al.\cite{telari2002Ptcharacterization} have demonstrated that deposited material of higher C\% and O\% are more restive than that of lower impurity contents. The higher resistivity of the $\mathrm{Rb^+}$ FIBID Pt also agrees with its faster deposition rates than $\mathrm{Ga^+}$ FIBID Pt, as Rue et al.\cite{rue2012XeFIBID} state that the resistivity grows with the deposition rate.

\subsection{\label{sec:surface}Deposit surface}

\begin{figure*}
\centering
\includegraphics[width=0.9\linewidth]{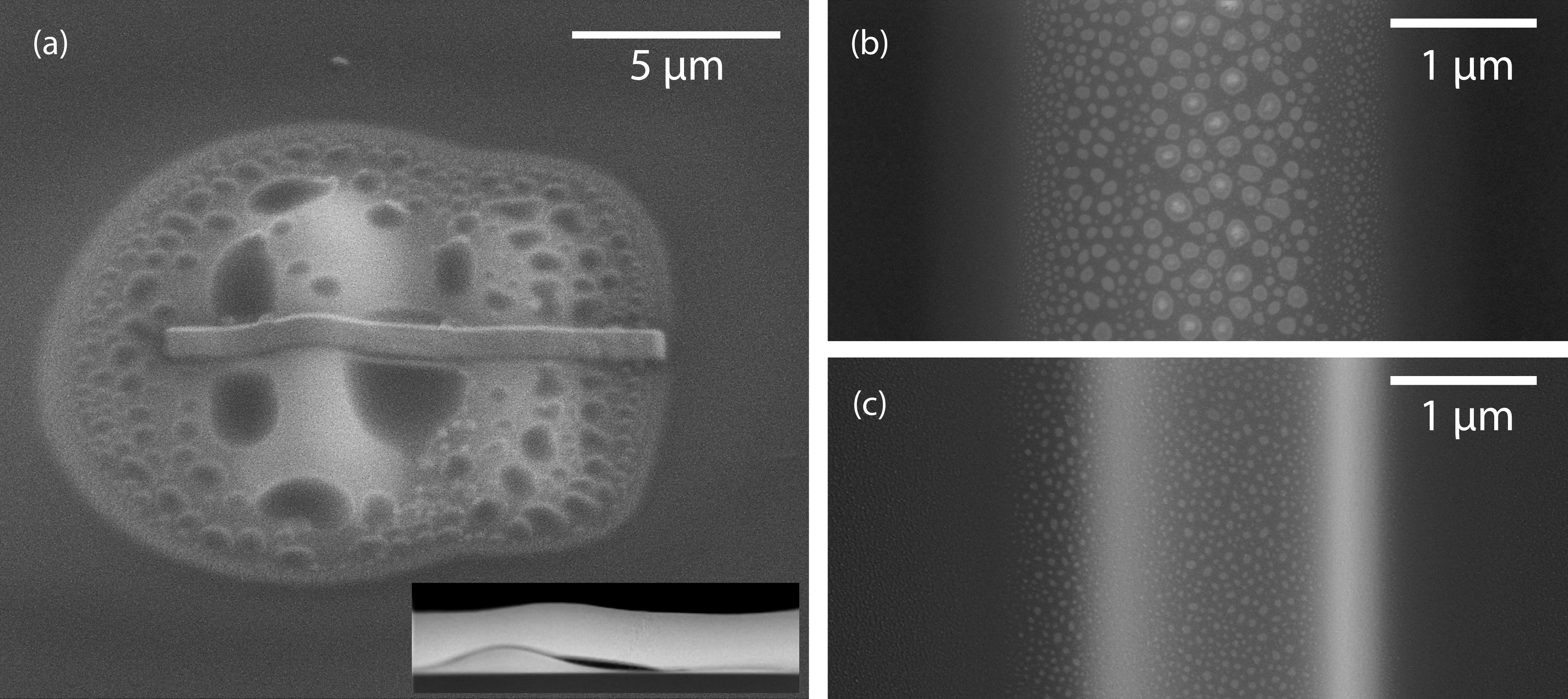}
\caption{Surface of $\mathrm{Rb^+}$ FIBID Pt (a) two $1.5\times45$ $\mu$m$^2$ FIBID Pt side by side created by a 8.5 keV 7.0 pA beam, left bar dose = $5.5\times10^{17}$ ions/cm$^2$ and right bar dose = $9.1\times10^{16}$ ions/cm$^2$. A Pt deposition protective cap lies over the region of interest selected for the TEM lamella sample. The inset shows the final lamella imaged by the TEM. (b) surface of a $\mathrm{Rb^+}$ FIBID Pt with ion dose = $9.1\times10^{16}$ ions/cm$^2$ (c) surface of a $\mathrm{Rb^+}$ FIBID Pt with ion dose = $5.5\times10^{17}$ ions/cm$^2$}
\label{fig:surface}
\end{figure*}

Fig.\ref{fig:surface}(a) and (b) shows bubble-like structures on the surface of the $\mathrm{Rb^+}$ FIBID Pt samples. As Fig.\ref{fig:surface}(b) and (c) demonstrate, these bubbles were more prevalent on deposits made with lower ion doses. With both deposits made by 6.5 pA $\mathrm{Rb^+}$, the deposit in Fig.\ref{fig:surface}(b) received about 1/5 of the ion dose as the one shown in Fig.\ref{fig:surface}(c). A similar effect is also shown between the left and right deposit bar in Fig.\ref{fig:surface}(a). The higher-dose left deposit contains far fewer bubbles on its surface than the lower-dose right bar. Bubbles could also be seen in the periphery of the deposits. These were believed to be caused by the beam halo around the central beam. The beam halo creates overspray deposit outside of the defined deposition area. With a much smaller current in the halo than in the central beam, the overspray deposits are equivalent to deposits made with low ion doses. 

TEM-EDS was used to study the chemical composition of these bubbles. The inset of Fig.\ref{fig:surface}(a) shows a lamella including both the high and low dose $\mathrm{Rb^+}$ FIBID Pt with the large bubble in the middle. The EDS results from this lamella reveal the bubble contains C:O:Pt:Rb = $15_{\pm2}$:$49_{\pm2}$:$20_{\pm1}$:$16_{\pm1}$ in atomic percentage, which has much lower C\% and Pt\% and much higher O\% compared to the bulk FIBID Pt. 

One speculation for the bubble formation process is that they are created by the $\mathrm{Rb^+}$ interacting with the volatile elements in the precursor gas, even though the exact mechanism is unknown. During the low-dose deposition, bonds of the precursor gas molecules were not as thoroughly broken by the ion-sample interaction as during the high ion dose. This lead to more volatile elements being trapped in the deposited material. Nonetheless, further study is required to fully understand the bubble forming process. It is worth noticing that these bubbles were only found on the surface and in the overspray of the FIBID Pt (as shown in the cross section Fig.\ref{fig:resistivity measurement}(b) so that they should not have interfered with the resistivity measurements. 

\section{\label{sec:conclusion}Conclusion}

Table.\ref{tab:FIBID properties} lists a summary of the properties of $\mathrm{Rb^+}$ and $\mathrm{Ga^+}$ FIBID Pt. Under similar beam conditions, both ion species can deposit Pt with a comparable rate and Pt purity near 40\%. The FIBID Pt by $\mathrm{Rb^+}$ contains slightly higher Pt\% than due to a lower Rb\% than Ga\%. However, the higher impurity contents such as C and O in the $\mathrm{Rb^+}$ FIBID Pt lead to higher resistivity of FIBID Pt by $\mathrm{Rb^+}$ than by $\mathrm{Ga^+}$. Lower dose $\mathrm{Rb^+}$ FIBID Pt is also susceptible to bubble formation on the surface of the deposited material. This effect can be reduced by increasing the ion dose, which will inevitably result in a thicker layer of deposit. Overall, $\mathrm{Rb^+}$ is a potential candidate as an alternative to $\mathrm{Ga^+}$ for FIBID at low beam energy and current for similar deposition results.

\begin{acknowledgments}
This work is part of the project Next-Generation Focused Ion Beam (NWO-TTW16178) of the research programme Applied and Engineering Sciences (TTW) which is (partly) financed by the Dutch Research Council (NWO). The authors are (partly) members of the FIT4NANO COST Action CA19140. The authors would like to acknowledge technical and equipment support by Kees Flipse, Erik Kieft ,and Dustin Laur as well as helpful discussions with Greg Schwind, Chad Rue, Yuval Greenzweig, Lisa McElwee-White, and Thomas Löber.
\end{acknowledgments}

\section*{Data Availability Statement}
The data that support the findings of this study are available from the corresponding author upon reasonable request.

\appendix

\begin{table*}
\caption{\label{tab:FIBID properties}Summary of FIBID Pt properties}
\begin{ruledtabular}
\begin{tabular}{ccccccc}
\textbf{Ion} & \textbf{E $\mathrm{(keV)}$}  & \textbf{I} \text{(pA)}  &\textbf{$\mathbbm{R}_{dep} \mathrm{(\mu m^3/nC)}$} &\textbf{Composition $\mathrm{(at \:\%)}$}\footnote{The atomic percentage results have 3-5\% statistical uncertainty.} &\textbf{$ \rho \mathrm{(\mu\Omega\cdot cm)}$}\footnote{10 pA current was used for making the $\mathrm{Rb^+}$ FIBID Pt sample and 7.8 pA current was used for the $\mathrm{Ga^+}$ FIBID Pt.}  &\textbf{Grain $\mathrm{(nm)}$}\\ \hline 
    \textbf{Rb\textsuperscript{+}}& 8.5    & 7.0   & $0.90_{\pm0.02}$   &C:O:Pt:Rb=25:20:49:5 &$1.2_{\pm0.4}\times 10^5$ &$13_{\pm3}$\\ 
    \textbf{Ga\textsuperscript{+}}           & 8.0      & 8.5      & $0.73_{\pm0.02}$ &C:O:Pt:Ga=22:14:37:27 &$6.4_{\pm0.7}\times 10^3$ & $14_{\pm2}$
\end{tabular}
\end{ruledtabular}
\end{table*}
\vspace{7.0pt}

\nocite{*}
\providecommand{\noopsort}[1]{}\providecommand{\singleletter}[1]{#1}%

\end{document}